\RequirePackage{fix-cm}
\documentclass[smallextended]{svjour3}       
\smartqed  
\usepackage{graphicx}
\usepackage{amsmath}
\usepackage{amssymb}
\begin{document}

\title{Neural News Recommendation with Negative Feedback}


\author{Chuhan Wu       \and
        Fangzhao Wu     \and
        Yongfeng Huang  \and
        Xing Xie
        }


\institute{Chuhan Wu, Yongfeng Huang \at
              Department of Electronic Engineering \& BNRist, Tsinghua University, Beijing 100084, China \\
              \email{wuchuhan15@gmail.com, yfhuang@tsinghua.edu.cn}           
           \and
           Fangzhao Wu, Xing Xie \at  
              Microsoft Research Asia, Beijing 100080, China \\
              \email{wufangzhao@gmail.com, xingx@microsoft.com}    
}

\date{Received: date / Accepted: date}

\maketitle

\begin{abstract}
News recommendation is important for online news services.
Precise user interest modeling is critical for personalized news recommendation.
Existing news recommendation methods usually rely on the implicit feedback of users like news clicks to model user interest.
However, news click may not necessarily reflect user interests because users may click a news due to the attraction of its title but feel disappointed at its content.
The dwell time of news reading is an important clue for user interest modeling, since short reading dwell time usually indicates low and even negative interest.
Thus, incorporating the negative feedback inferred from the dwell time of news reading can improve the quality of user modeling.
In this paper, we propose a neural news recommendation approach which can incorporate the implicit negative user feedback.
We propose to distinguish positive and negative news clicks according to their reading dwell time, and respectively learn user representations from positive and negative news clicks via a combination of Transformer and additive attention network. 
In addition, we propose to compute a positive click score and a negative click score based on the relevance between candidate news representations and the user representations learned from the positive and negative news clicks.
The final click score is a combination of positive and negative click scores.
Besides, we propose an interactive news modeling method to consider the relatedness between title and body in news modeling.
Extensive experiments on real-world dataset validate that our approach can achieve more accurate user interest modeling for news recommendation.

\end{abstract}

\section{Introduction}

Online news services such as Google News\footnote{https://news.google.com} and Microsoft News\footnote{https://www.msn.com} can collect news from various sources and display them to users in a unified view~\cite{das2007google,wu2020mind}.
However, a large number of news articles are generated every day and it is overwhelming for users to find their interested news~\cite{okura2017embedding}.
Thus, personalized news recommendation is critical for these news services to target user interests and alleviate information overload~\cite{wu2019neural}.

Accurate user interest modeling is a core problem in news recommendation.
Existing news recommendation methods usually learn representations of users from their historical news click behaviors~\cite{okura2017embedding,wang2018dkn,wu2019npa,wu2019neuralnrms}.
For example, Okura et al.~\cite{okura2017embedding} proposed to learn user representations from the clicked news via a gated recurrent unit (GRU) network.
Wang et al.~\cite{wang2018dkn} proposed to learn user representations from clicked news with a candidate-aware attention network that evaluates the relevance between clicked news and candidate news.
In fact, users usually click news due to their interests in the news titles.
However, news titles are usually short and the information they condensed is very limited and even misleading. 
Users may be disappointed after reading the content of their clicked news.
Thus, news click behaviors do not necessarily indicate user interest, and modeling users solely based on news clicks may not be accurate enough.

\begin{figure}[!t]
  \centering
    \includegraphics[width=0.98\linewidth]{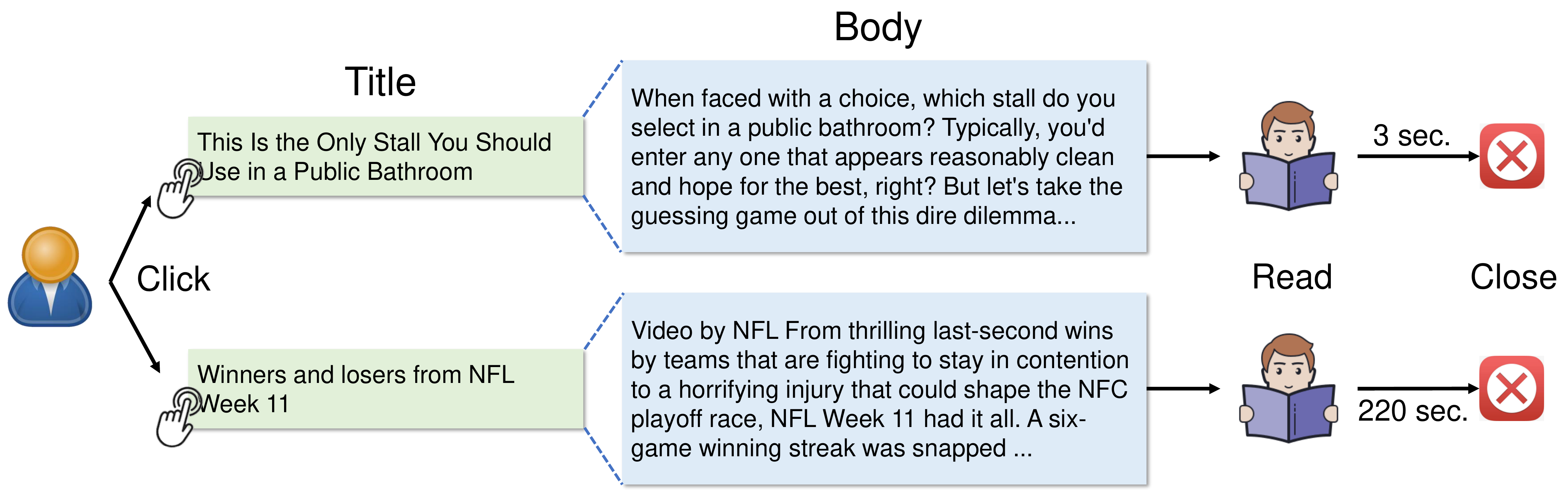}
  \caption{Two news clicked by a user. The time represent the news reading dwell time.}
  \label{fig.example}
\end{figure}

Besides news clicks, users also provide other implicit feedback like the dwell time of news reading, which is an important indication of user interest.
As shown in Fig.~\ref{fig.example}, the user clicks two news articles and respectively reads them for 3 and 220 seconds before closing the news webpages.
We can infer that this user may not be interested in the first news because she reads the body of this news for only 3 seconds before her leaving.
Thus, a short dwell time of news reading can be regarded as a kind of implicit negative user feedback, which can help model user interest more accurately.
In addition, since users usually decide which news to click according to news titles and read news bodies for detailed information, incorporating both news title and body is useful for news and user interest modeling.
Moreover, the title and body of the same news usually have some inherent relatedness in describing news content.
For example, in Fig.~\ref{fig.example} the title of the second news indicates that this news is about NFL games, and the body introduces the details of these events.
Capturing the relations between news title and body may help better understand news content and facilitate subsequent user interest modeling.

In this paper, we propose a neural news recommendation approach with negative feedback (NRNF).
We propose to distinguish positive and negative news click behaviors according to the reading dwell time, and learn separate user representations from the positive and negative news clicks with a Transformer to capture behavior relations and an additive attention network to select important news for user modeling.
In addition, we propose to respectively compute a positive click score and a negative click score based on the relevance between candidate news representation and the user representations learned from positive and negative news clicks.
Besides, we propose an interactive news modeling method to incorporate both news title and body by using an interactive attention network to model their relatedness.
Massive experiments conducted on real-world dataset demonstrate that our approach can effectively improve the performance of user modeling by incorporating implicit negative user feedback.

\section{Related Work}\label{sec:RelatedWork}

News recommendation is an important task in the data mining field, and has gained increasing attention in recent years~\cite{zheng2018drn}.
Learning accurate representations of users is critical for news recommendation.
Many existing news recommendation methods rely on manual feature engineering to build user representations~\cite{liu2010personalized,capelle2012semantics,son2013location,karkali2013match,garcin2013personalized,bansal2015content,ren2015personalized,chen2017location,zihayat2019utility}.
For example, Li et al.~\cite{li2010contextual} proposed  to  represent users using their demographics, geographic information and behavioral categories which summarize the consumption history on Yahoo!.
Garcin et al.~\cite{garcin2012personalized} proposed to model users by aggregating the LDA features of all clicked news into a user vector by averaging.
Lian et al.~\cite{lian2018towards} proposed to build user representations from various handcrafted features like user ID, demographics, locations and the feature collection of clicked news.
However, these methods rely on manual feature engineering, which needs massive domain knowledge to craft.
In addition, handcrafted features are usually not optimal for representing user interest.

In recent years, several news recommendation methods based on deep learning techniques are proposed~\cite{okura2017embedding,khattar2018weave,wang2018dkn,wu2019neural,an2019neural,wu2019neuralnaml,wu2019neuralnrhub,wu2019neuralnrms,ge2020graph,hu2020graph}.
For example, Okura et al.~\cite{okura2017embedding} proposed to learn representations of users from the representations of their browsed news using a GRU network.
Wang et al.~\cite{wang2018dkn} proposed to learn user representations from clicked news based on their relevance to the candidate news.
Zhu et al.~\cite{zhu2019dan} proposed to generate user representations using a combination of a CNN network and an attention-based LSTM network.
Wu et al.~\cite{wu2019neuralnrms} proposed to use multi-head self-attention network to learn user representations by capturing the interactions between clicked news.
These methods only rely on news click behaviors to learn user representations, which may be insufficient because news clicks may not exactly reflect user preferences.
Different from these methods, our approach can learn representations of users by incorporating the implicit negative feedback of users inferred from their reading dwell times, which can provide useful clues to calibrate the  user interest inferred from news click behaviors.

\section{Our Approach}\label{sec:Model}

\begin{figure*}[!t]
  \centering
    \includegraphics[width=0.99\linewidth]{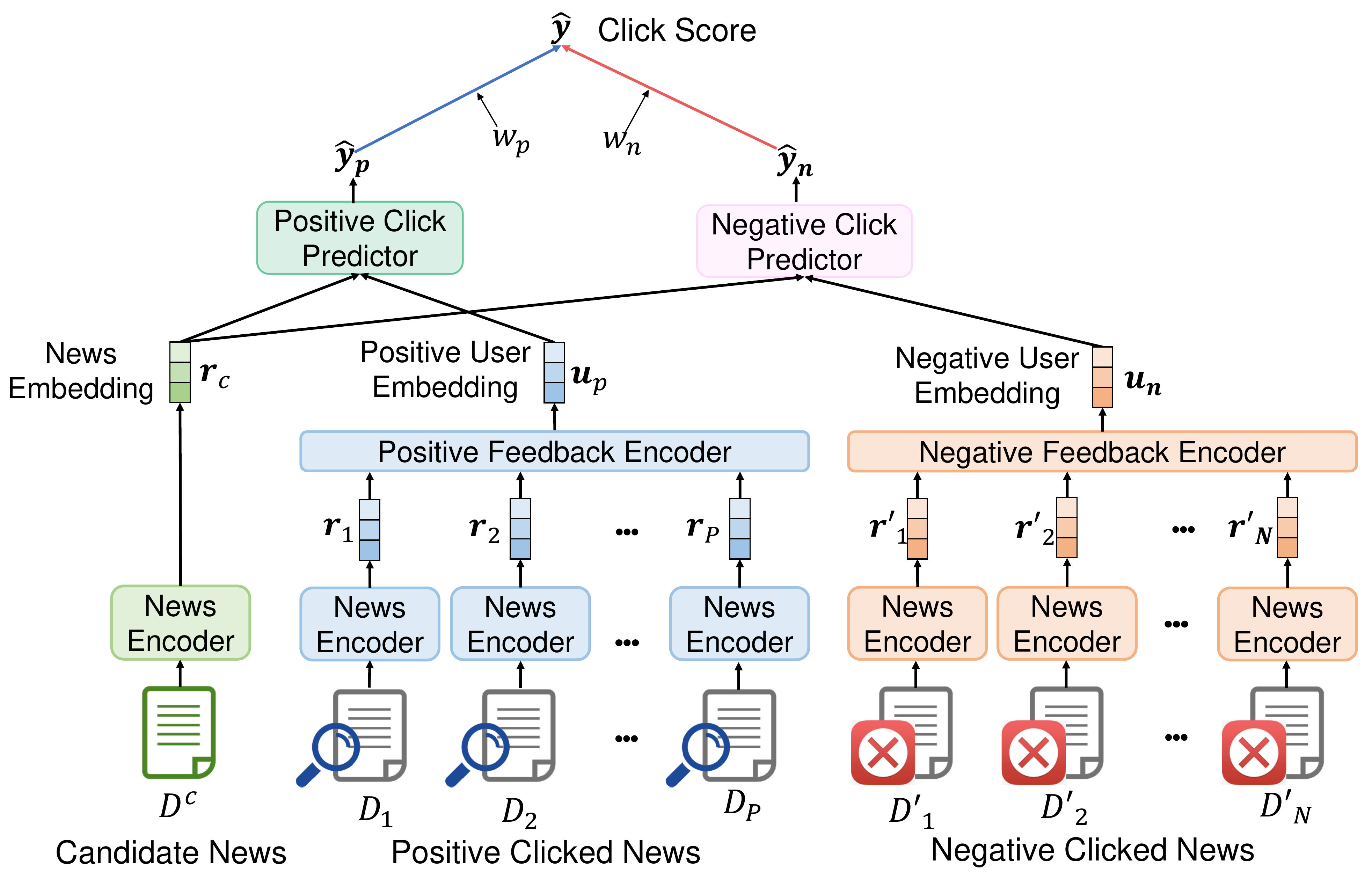}

  \caption{The framework of our \textit{NRNF} approach.}
  \label{fig.model}
\end{figure*}

In this section, we introduce the details of our neural news recommendation approach with negative feedback (NRNF).
The architecture of \textit{NRNF} is illustrated in Fig.~\ref{fig.model}.
We will introduce each core technique in our method, including user modeling, click prediction, news modeling, and model training, in the following sections.

\subsection{User Modeling with Negative Feedback}

In this section, we introduce how to model user interests from their clicked news by considering negative feedback.
Usually, the interests of users on the news they read carefully are very different from those they closed  quickly after having a glance at the body. 
For example, as shown in Fig.~\ref{fig.example}, we can infer that the user may be indeed interested in the second news since she reads this news for about 220 seconds.
However, she is probably not interested in or unsatisfied with the first news since she reads this news only for 3 seconds after click.
Thus, incorporating short reading dwell time as the implicit negative feedback of users has the potential to enhance the modeling of user interest.
To characterize the differences of news in representing the real user interests, we use their reading dwell times $t$ to distinguish positive news clicks from the negative ones.
We regard the clicked news with $t<T$ as negative news clicks and those with  $t\geq T$ as positive news clicks, where $T$ is a time threshold.
Since simply aggregating all news together will ignore the differences of positive and negative news, in the \textit{user encoder} module we propose to use a multi-view learning framework to incorporate positive and negative news clicks as different views of users.
Their details are introduced as follows.

The positive news click view is used to learn a positive user embedding  from positive clicked news.
We denote the $P$ positive clicked news of a user as $[D_1, D_2, ..., D_P]$.
We first use a news encoder to obtain their representations, which are denoted as $\mathbf{R}^p=[\mathbf{r}_1, \mathbf{r}_2, ..., \mathbf{r}_P]$.
Then, we use a positive feedback encoder to learn a unified user embedding from these news representations.
It uses a similar architecture with the user encoder used by Wu et al.~\cite{wu2019neuralnrms}.
Since different clicked news usually have some interactions with each other, we use a news-level Transformer~\cite{vaswani2017attention} to capture the interactions among them\footnote{We add position embeddings to the news embeddings to capture behavior orders.}, and the output sequence is computed as:
\begin{equation}
    \mathbf{D}^p=[\mathbf{d}_1,\mathbf{d}_2,...,\mathbf{d}_P]=Transformer(\mathbf{R}^p,\mathbf{R}^p,\mathbf{R}^p),
\end{equation}
where the three parameters in $Transformer(\cdot)$ respectively denote the input query, key and value. 
In addition, since different clicked news have different importance in modeling user interests, we use an additive attention~\cite{yang2016hierarchical} network to select important news to learn more informative user representations.
The final positive user embedding $\mathbf{u}_p$ learned from this view is computed as:
\begin{equation}
\mathbf{u}_p=Attention(\mathbf{q}^p,\mathbf{D}^p,\mathbf{D}^p),
\end{equation}
where $\mathbf{q}^p$ is a query parameter vector, $\mathbf{D}^p$ is regarded as the key and value. 
It summarizes user interests inferred from the positive user feedback.

The negative news click view is used to learn representations of users from their negative clicked news.
Similar to positive news, we use a combination of Transformer and additive attention network to learn the negative user embedding from the representations of the negative clicked news.
We denote the negative user embedding learned from this view as  $\textbf{u}_n$, which condenses the information of negative user interest  inferred from the negative clicked news.

\subsection{Click Prediction}

In this section, we introduce the details of click prediction in our approach.
Usually, if a candidate news is similar to the news that a user reads carefully, it may have a high probability to be clicked by this user.
However, if it is very similar to the news that are previously closed by a user very quickly, it will probably not be clicked by this user.
Motivated by the observations above, we propose to calculate a positive click score $\hat{y}_p$ and a negative click score $\hat{y}_n$ respectively based on the relevance of candidate news to the positive and negative user interest. 
We denote the embedding of a candidate news $D^c$ as $\mathbf{r}_c$.
Following~\cite{okura2017embedding}, $\hat{y}_p$ and $\hat{y}_n$ are respectively computed by the inner product of the candidate news embedding  and the user embeddings learned from the positive and negative views, which are formulated as:
\begin{equation}
\begin{split}
\hat{y}_p & =  \mathbf{r}_c^T\mathbf{u}_p,\\
\hat{y}_n & = \mathbf{r}_c^T\mathbf{u}_n.
\end{split}
\end{equation}
The final click probability score $\hat{y}$ is a weighted summation of the positive score $\hat{y}_p$ and the negative score $\hat{y}_n$, which is formulated as:
\begin{equation}
\hat{y} = w_p\hat{y}_p+w_n\hat{y}_n,
\end{equation} 
where $w_p$ and $w_n$ are learnable parameters.
A positive parameter means that the relevance between candidate news and the corresponding user embedding has a positive correlation with the final click score, and this correlation is negative if the parameter is negative.

\subsection{News Modeling}

In this section, we introduce the architectures of the news encoder in our approach for news modeling.
It is used to learn representations of news from their titles and bodies.
Since news titles and bodies usually have very different characteristics, simply concatenating title and body together may not be optimal.
Thus, in this module we use a multi-view learning framework to incorporate the information of title and body as different views of news.
In addition, since there is usually inherent relatedness between news title and body, we propose to use an interactive attention network to capture their relations and form the unified news embedding. 
The details of each module in the news encoder are introduced as follows.

\begin{figure*}[!t]
  \centering
    \includegraphics[width=0.7\linewidth]{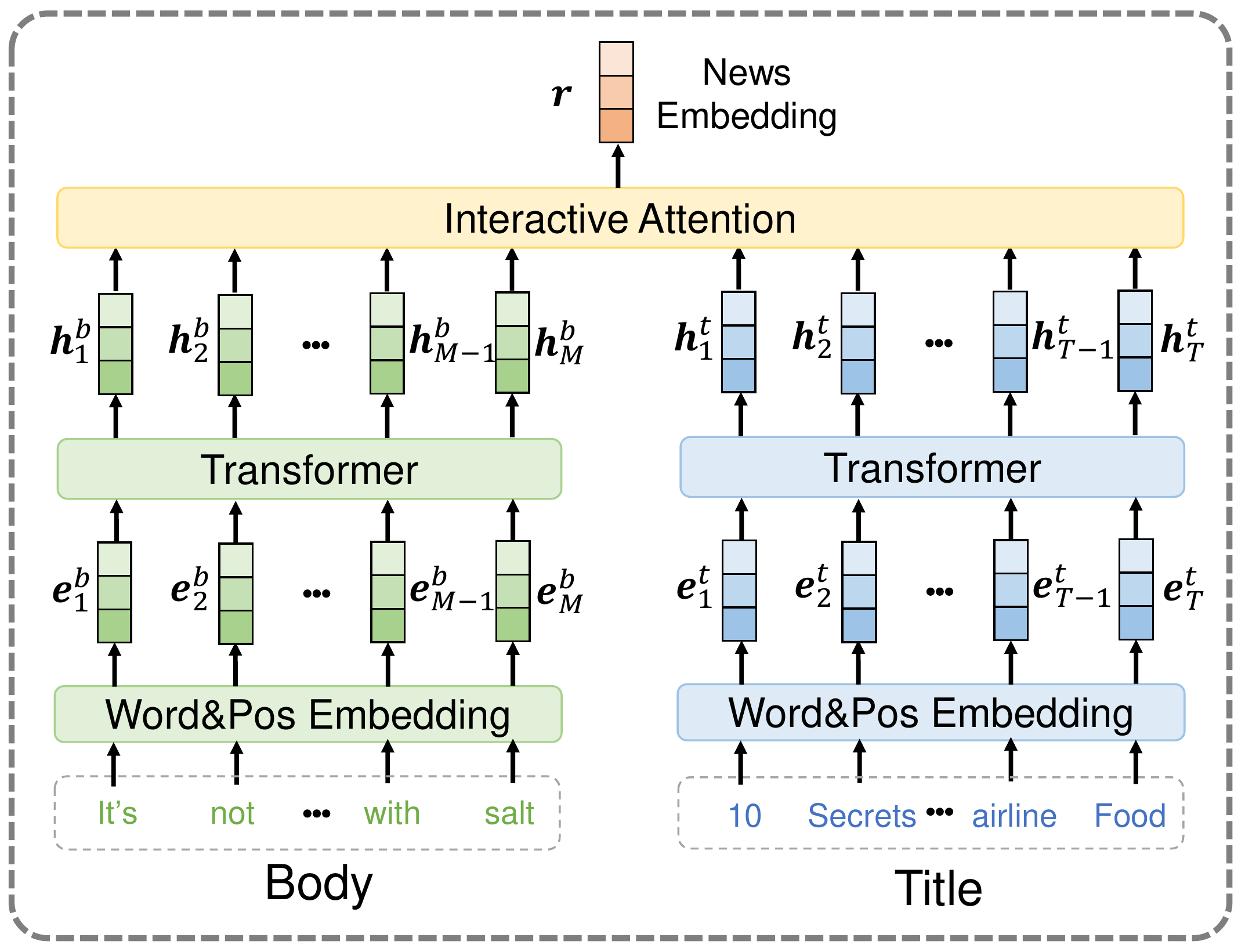}

  \caption{The architecture of the news encoder.}
  \label{fig.model2}
\end{figure*}

The first module in the news encoder is a title transformer, which is used to capture the contexts of words in a news title.
We first use a word and position embedding layer to convert the sequence of words in a news title into a sequence of low-dimensional word embeddings with their position information.
Denote a news title with $T$ words as $[w^t_1,w^t_2,...,w^t_T]$.
It is converted into an embedding sequence $\mathbf{E}^t=[\mathbf{e}^t_1,\mathbf{e}^t_2,...,\mathbf{e}^t_T]$.
Then, we use a Transformer to learn the contextual representation of each word, which is computed as follows: 
\begin{equation}
\mathbf{H}^t=[\mathbf{h}^t_1,\mathbf{h}^t_2,...,\mathbf{h}^t_T]=Transformer(\mathbf{E}^t,\mathbf{E}^t,\mathbf{E}^t).
\end{equation}

The second module in the news encoder is a body transformer, which aims to capture the contexts of words in a news body.
Similar to news title, we also use a word and position embedding layer to obtain the embeddings of the $M$ words in a news body, which are denoted as $\mathbf{E}^b=[\mathbf{e}^b_1,\mathbf{e}^b_2,...,\mathbf{e}^b_M]$. 
Then, we use a Transformer to capture the contexts of words, and its output  $\mathbf{H}^b$ is computed as 
\begin{equation}
\mathbf{H}^b=[\mathbf{h}^b_1,\mathbf{h}^b_2,...,\mathbf{h}^b_M]=Transformer(\mathbf{E}^b,\mathbf{E}^b,\mathbf{E}^b).
\end{equation}

The third module is an interactive attention network, which is used to capture the relations between news title and body when forming the unified news embedding. 
We first use two additive attention network to respectively attend to the important words in news title and body to form the title and body representations(denoted as $\mathbf{r}^t$ and $\mathbf{r}^b$), which are formulated as  
\begin{equation}
    \mathbf{r}^t=Attention(\mathbf{q}^t,\mathbf{H}^t,\mathbf{H}^t),
\end{equation}
\begin{equation}
\mathbf{r}^b=Attention(\mathbf{q}^b,\mathbf{H}^b,\mathbf{H}^b),
\end{equation}
where $\mathbf{q}^t$ and $\mathbf{q}^b$ are learnable query vectors.
Then, we use the title representation as the query of the body attention network to select words in news body according to their relevance to the content of news title, and the output title-aware body representation $\mathbf{c}^{b}$ is computed as follows:
\begin{equation}
\mathbf{c}^{b}=Attention(\mathbf{r}^t,\mathbf{H}^b,\mathbf{H}^b).
\end{equation}
We learn a body-aware title representation $\mathbf{c}^{t}$  in a similar way by using the body representation as the title attention query, which is formulated as:
\begin{equation}
\mathbf{c}^{t}=Attention(\mathbf{r}^t,\mathbf{H}^b,\mathbf{H}^b).
\end{equation}
We finally aggregate all title and body representations into a unified news representation $\mathbf{r}$, i.e., $\mathbf{r}=\mathbf{r}^t+\mathbf{r}^b+\mathbf{c}^t+\mathbf{c}^b$.

\subsection{Model Training}

Following~\cite{huang2013learning,wu2019npa}, we use negative sampling techniques to construct labeled data from the raw news impression logs for model training.
For each news clicked by a user, we randomly sample $Q$ news that are displayed in the same impression but not clicked by this user.
We denote the click probability scores of the clicked news sample as $\hat{y}^+$ and the $Q$ unclicked news samples as $[\hat{y}_1^-, \hat{y}_2^-, ...,\hat{y}_Q^-]$.
we re-formulate the news click prediction problem as a pseudo $Q+1$-way classification task to jointly predict these scores.
We normalize these scores via the softmax function to compute the posterior click probability of the clicked news, which is formulated as follows:
\begin{equation}
p_i=\frac{\exp(\hat{y}_i^+)}{\exp(\hat{y}_i^+)+\sum_{j=1}^{Q}{\exp(\hat{y}_{i,j}^-)}}.
\end{equation}
The final loss function we used is the negative log-likelihood of all clicked news samples, which is formulated as:
\begin{equation}
 \mathcal{L}=-\sum_{i\in \mathcal{S}}\log(p_i),
\end{equation}
where $\mathcal{S}$ is the set of clicked news samples for training.

\section{Experiments}\label{sec:Experiments}

\subsection{Datasets and Experimental Settings}

Our experiments were conducted on a real-world news recommendation dataset collected from MSN News\footnote{https://www.msn.com/en-us/news} logs  from Oct. 31, 2018 to Jan. 29, 2019.
The detailed statistics are shown in Table~\ref{dataset}.
We sorted all sessions by time, and we used the first 80\% of sessions for training, the next 10\% for validation and the rest 10\% for test.

\begin{table}[h]

\caption{Statistics of the dataset.}

\begin{tabular}{|l|r|l|r|}
\hline
\textbf{\# users}       & 337,699     & \textbf{avg. \# words per title} & 10.99      \\ 
\textbf{\# news}        & 249,038      & \textbf{avg. \# words per body} & 719.34      \\ 
\textbf{\# sessions} & 500,000     & \textbf{avg. dwell time}    & 294.91 s\\ 
\textbf{\# samples} & 40,400,206  & \textbf{avg. \# negative news per user}     & 5.00  \\ 
\textbf{\# clicked samples}     & 990,624     & \textbf{\# unclicked samples}     & 39,409,582    \\ \hline
\end{tabular}

\label{dataset}
\end{table}

In our experiments, the word embeddings were 300-dimensional and initialized by Glove~\cite{pennington2014glove}
The Transformers had 8 attention heads, whose output dimension was 32.
The maximum lengths of title and body were respectively set to 30 and 200, respectively.
The numbers of positive and negative clicked news were respectively limited to 50 and 20.
Following~\cite{huang2013learning}, the negative sampling ratio $Q$ was set to 4.
The threshold $T$ was set to 10 seconds.
Adam~\cite{kingma2014adam} was used as the optimizer, and the batch size was 30.
To mitigate overfitting, we applied dropout~\cite{srivastava2014dropout} techniques to the word embeddings and Transformer networks, and the dropout ratio was set to 0.2.
These hyperparameters were tuned on the validation set.
We independently repeated each experiment 10 times and reported the average results in terms of AUC, MRR, nDCG@5 and nDCG@10.

\subsection{Performance Evaluation}
We evaluate the performance of our approach by comparing it with many baseline methods, including:
\begin{itemize}
\item \textit{LibFM}~\cite{rendle2012factorization}, a factorization machine based method for recommendation. We extracted the TF-IDF features from the titles and bodies of news clicked by users to build their representations.
\item  \textit{DSSM}~\cite{huang2013learning}, deep structured semantic model. We regarded the concatenation of the news clicked by a user as the query and candidate news as documents, and we used the same feature with \textit{LibFM} for news and user representation.
\item  \textit{Wide\&Deep}~\cite{cheng2016wide}, a famous neural recommendation method which consists of a wide linear channel and a deep neural network channel. We used the same features with \textit{LibFM} as the input of both channels.
\item  \textit{DeepFM}~\cite{guo2017deepfm}, another widely used neural recommendation method which uses a combination of factorization machines and deep neural networks. We used the same TF-IDF features to feed for both components.
\item  \textit{DFM}~\cite{lian2018towards}, a deep fusion model for news recommendation, which combines fully connected layers with different depths. We used TF-IDF features and word embeddings as the input.
\item  \textit{EBNR}~\cite{okura2017embedding}, an embedding based neural news recommendation method which uses denoising autoencoders to learn news representations and a GRU network to learn user representations.
\item  \textit{DKN}~\cite{wang2018dkn}, a neural news recommendation method which learns news representations via  knowledge-aware CNNs and learns user representations via an attention network based on the relevance between clicked news and candidate news. 
\item  \textit{DAN}~\cite{zhu2019dan}, a neural news recommendation method which learns news representations via two parallel CNNs and learns user representations using a combination of attentional LSTM and CNN;
\item  \textit{NAML}~\cite{wu2019neuralnaml}, a neural news recommendation approach with attentive multi-view learning.
\item \textit{NRMS}~\cite{wu2019neuralnrms}, a neural news recommendation method that uses multi-head self-attention for news and user modeling.
\item  \textit{NRNF-basic}, a variant of our \textit{NRNF} approach that does not consider negative feedback.
\item  \textit{NRNF}, our news recommendation approach with negative feedback.
\end{itemize}
For fair comparison, in \textit{GRU}, \textit{DKN}, \textit{DAN} and \textit{NRMS} we used the concatenation of news title and news body to learn news representations.
The results of these methods are summarized in Table~\ref{table.result}.

\begin{table*}[t]
	\centering

\caption{The performance of different methods. The improvement over other methods is significant at $p<0.01$.}\label{table.result}
		\resizebox{1\textwidth}{!}{
\begin{tabular}{|c|c|c|c|c|}
\hline
    \textbf{Methods}         & \textbf{AUC}             & \textbf{MRR}             & \textbf{nDCG@5}          & \textbf{nDCG@10}         \\ \hline
LibFM~\cite{rendle2012factorization}        & 0.6296$\pm$0.0049          & 0.2233$\pm$0.0041          & 0.2467$\pm$0.0045          & 0.3044$\pm$0.0039          \\
DSSM~\cite{huang2013learning}         & 0.6514$\pm$0.0031          & 0.2300$\pm$0.0028          & 0.2539$\pm$0.0032          & 0.3099$\pm$0.0030          \\ 
Wide\&Deep~\cite{cheng2016wide} & 0.6367$\pm$0.0029            & 0.2257$\pm$0.0026          & 0.2496$\pm$0.0024          & 0.3050$\pm$0.0022          \\
DeepFM~\cite{guo2017deepfm}       & 0.6410$\pm$0.0041          & 0.2275$\pm$0.0035          & 0.2502$\pm$0.0037          & 0.3069$\pm$0.0032          \\ 
DFM~\cite{lian2018towards}          & 0.6444$\pm$0.0037          & 0.2288$\pm$0.0033         & 0.2511$\pm$0.0038          & 0.3082$\pm$0.0035          \\ \hline
EBNR~\cite{okura2017embedding}          & 0.6554$\pm$0.0022          & 0.2324$\pm$0.0019          & 0.2556$\pm$0.0024          & 0.3125$\pm$0.0021         \\
DKN~\cite{wang2018dkn}          & 0.6531$\pm$0.0036          & 0.2311$\pm$0.0028         & 0.2544$\pm$0.0030          & 0.3112$\pm$0.0026          \\ 
DAN~\cite{zhu2019dan}          & 0.6621$\pm$0.0016& 0.2354$\pm$0.0012 & 0.2588$\pm$0.0014& 0.3153$\pm$0.0012 \\ 
NAML~\cite{wu2019neuralnaml}          & 0.6739$\pm$0.0015& 0.2412$\pm$0.0014 & 0.2634$\pm$0.0013& 0.3215$\pm$0.0016 \\ 
NRMS~\cite{wu2019neuralnrms}          & 0.6789$\pm$0.0019& 0.2440$\pm$0.0017 & 0.2671$\pm$0.0017& 0.3272$\pm$0.0018 \\ \hline
NRNF-basic          & \underline{0.6818}$\pm$0.0016& \underline{0.2456}$\pm$0.0016 & \underline{0.2688}$\pm$0.0017& \underline{0.3290}$\pm$0.0019 \\ 
NRNF*          & \textbf{0.6896}$\pm$0.0015& \textbf{0.2506}$\pm$0.0014 & \textbf{0.2780}$\pm$0.0015& \textbf{0.3323}$\pm$0.0016 \\ \hline
\end{tabular}
}
\end{table*}

From Table~\ref{table.result}, we have several observations.
First, compared with baseline methods that use handcrafted features to represent news and users (e.g., \textit{LibFM}, \textit{DSSM} and \textit{DeepFM}), the models using neural networks for learning news and user representations (e.g., \textit{EBNR}, \textit{NAML} and \textit{NRNF})  achieve better recommendation performance.
This is probably because neural networks can learn more informative news and user representations than hand-crafted features.
Second, compared with other neural news recommender like \textit{DAN}, \textit{NAML} and \textit{NRMS}, \textit{NRNF} and its variant \textit{NRNF-basic} achieve better performance.
This is because the latter methods can capture the relatedness between news title and body to enhance news modeling, while other methods cannot.
Third, \textit{NRNF} outperforms other baseline methods, and further hypothesis tests show that the improvement is significant.
This is probably because incorporating the implicit negative user feedback can help learn more accurate user interest representations.

\subsection{Effectiveness of Different News Information}

\begin{figure}[t]
	\centering
	\includegraphics[width=0.6\textwidth]{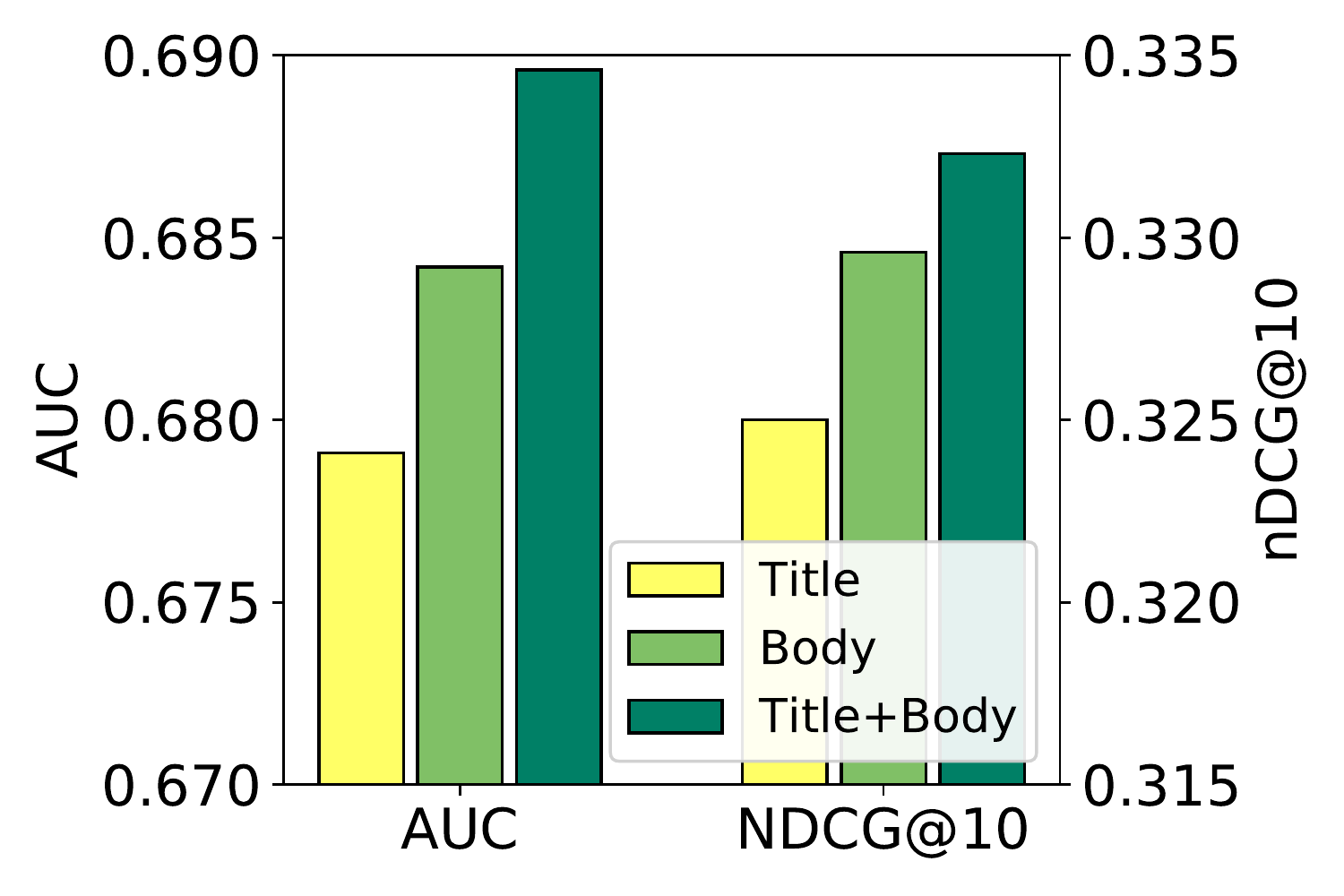}
\caption{Effectiveness of different news information.}\label{fig.view}
\end{figure}

Next, we explore the influence of incorporating different news information for news modeling.
We compare the performance of our approach with its variants with different combinations of news information, and the results are shown in Fig.~\ref{fig.view}.
From Fig.~\ref{fig.view}, we have several observations.
First, both news title and body are useful for news recommendation.
This is because users usually decide which news to click based on its title, and read news body for detailed information. 
Thus, both news title and body can provide important information for news recommendation.
Second, news body is more useful than news title in representing news.
This is intuitive because news body usually contains the details of news, which can provide richer information than news title.
Besides, combining both title and body of news can further improve the performance of our approach.
These results validate the effectiveness of the multi-view learning framework in our approach.  

\subsection{Effectiveness of Attention Network}

\begin{figure}[t]
	\centering

	\includegraphics[width=0.6\textwidth]{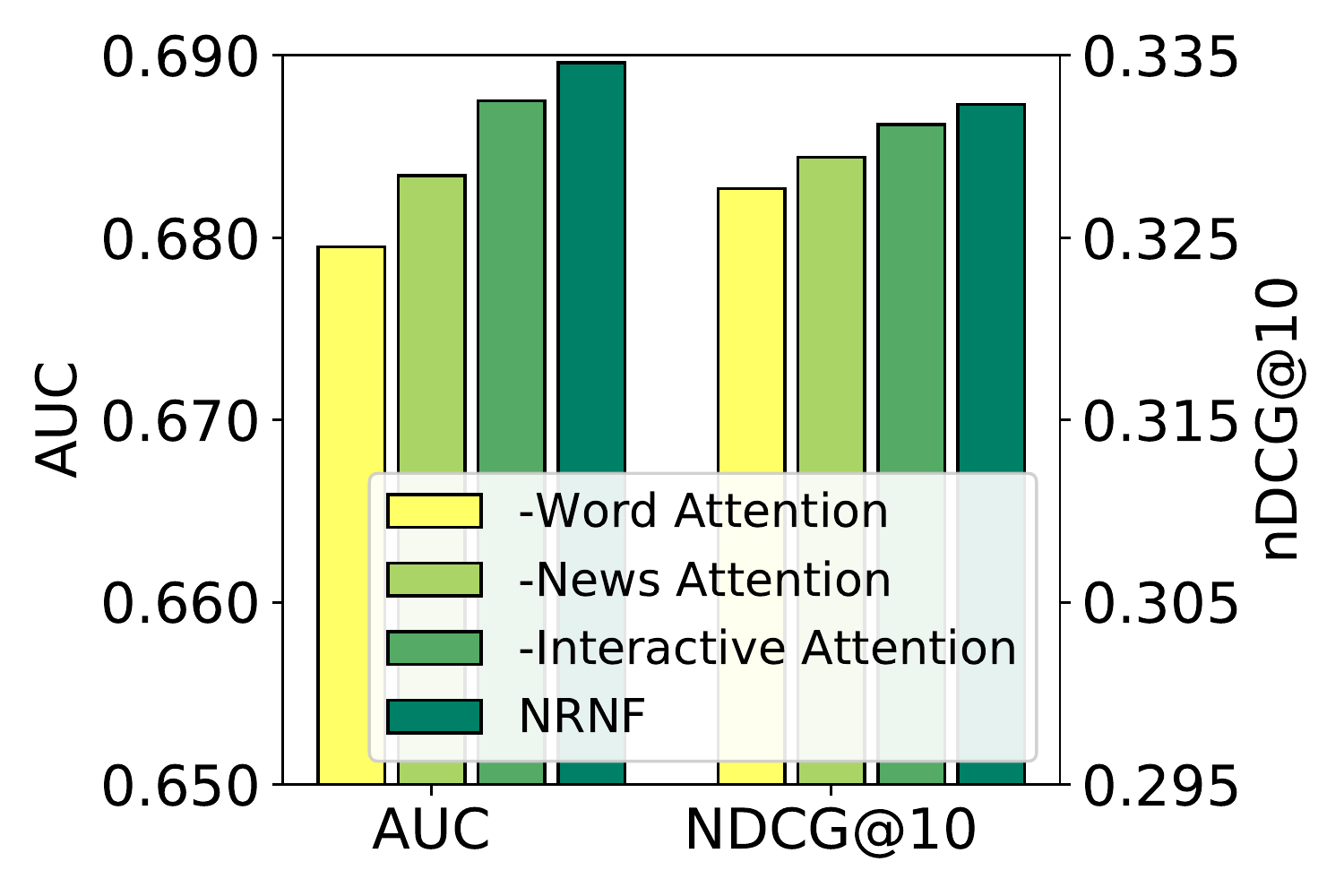}
	
\caption{Effectiveness of different attention networks.}\label{fig.att}
\end{figure}

Then, we explore the influence of different attention networks on the performance of our approach.
The results are shown in Fig.~\ref{fig.att}.
From Fig.~\ref{fig.att}, we find that word-level attention networks are very useful for our approach.
This is because different words in news titles and bodies usually have different importance in representing news.
Thus, selecting important words can help learn more informative news representations.
In addition, news-level attention network is also important.
This is because different news usually have different informativeness in representing the preferences of users, and selecting informative news can benefit user representation learning.
Besides, the interactive attention network can also boost the performance.
This may be because it can help capture the relations between news title and body to enhance news modeling.
Moreover, combining all these kinds of attention networks can further improve the performance of our approach.
These results verify the effectiveness of each kind of attention mechanism in our approach.

\subsection{Influence of Hyperparameters}
\begin{figure}[t]
	\centering

	\includegraphics[width=0.6\textwidth]{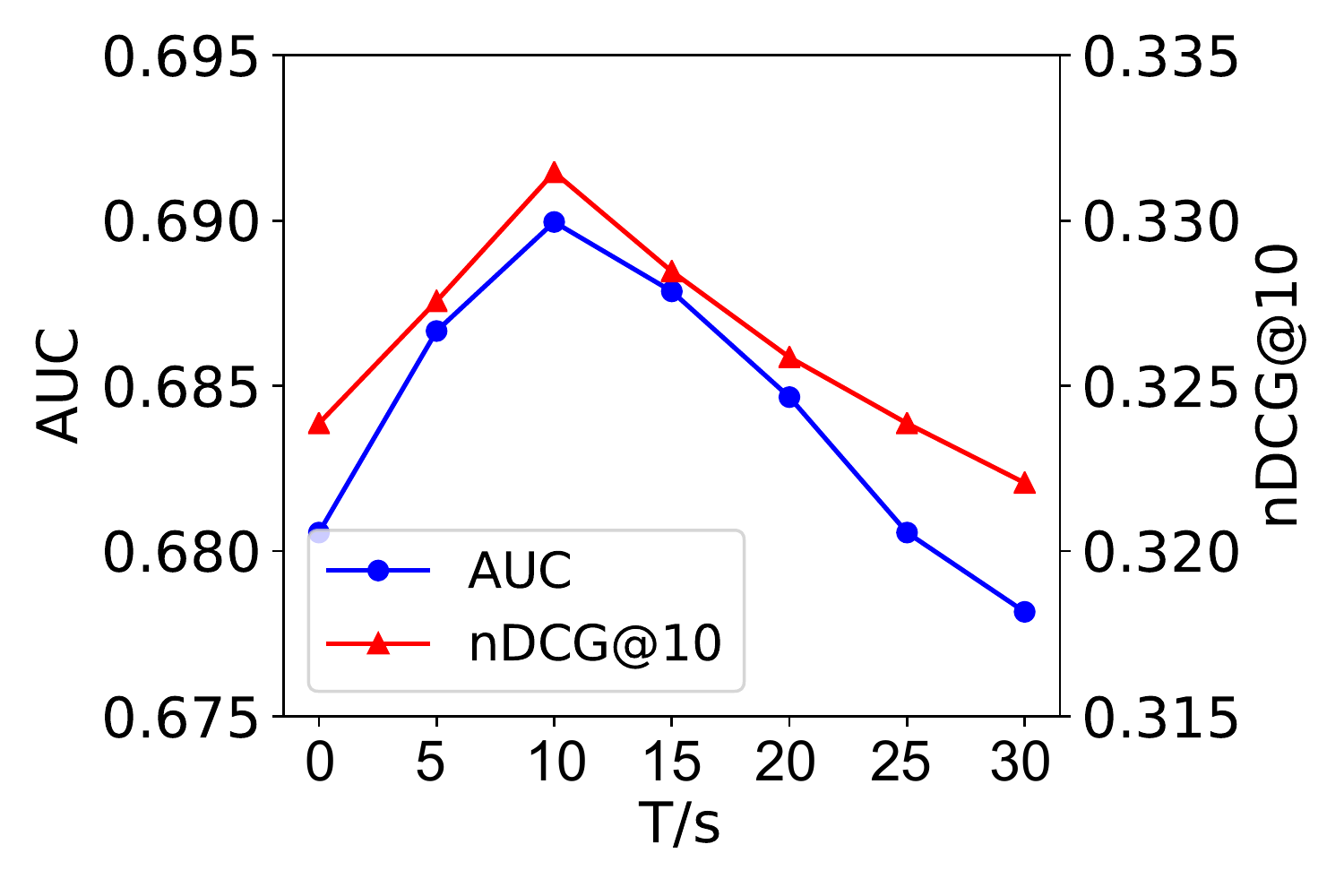}
	
\caption{Influence of the threshold $T$ on our approach.}\label{fig.th}
\end{figure}

In this section, we study the influence of an important hyperparameter on our approach, i.e., the dwell time threshold $T$.
We compare the performance of our approach w.r.t different values of $T$.
The results are illustrated in Fig.~\ref{fig.th}. 
We find the performance of our approach improves when the threshold $T$ increases.
This is probably because when $T$ is too small, many negative news clicks cannot be distinguished from the positive ones, and the modeling of user interest is not accurate enough.
In addition, the performance of our approach declines when $T$ goes too large.
This is also intuitive because when $T$ is too large, many positive news clicks will be mistakenly regarded as negative ones and the errors will be encoded into the negative user embedding, which leads to the sub-optimal  performance.
Therefore, a moderate selection of $T$, e.g., 10 seconds may be more appropriate for our approach, which is also consistent with our news browsing experiences. 

\begin{figure}[t]
	\centering

	\includegraphics[width=0.6\textwidth]{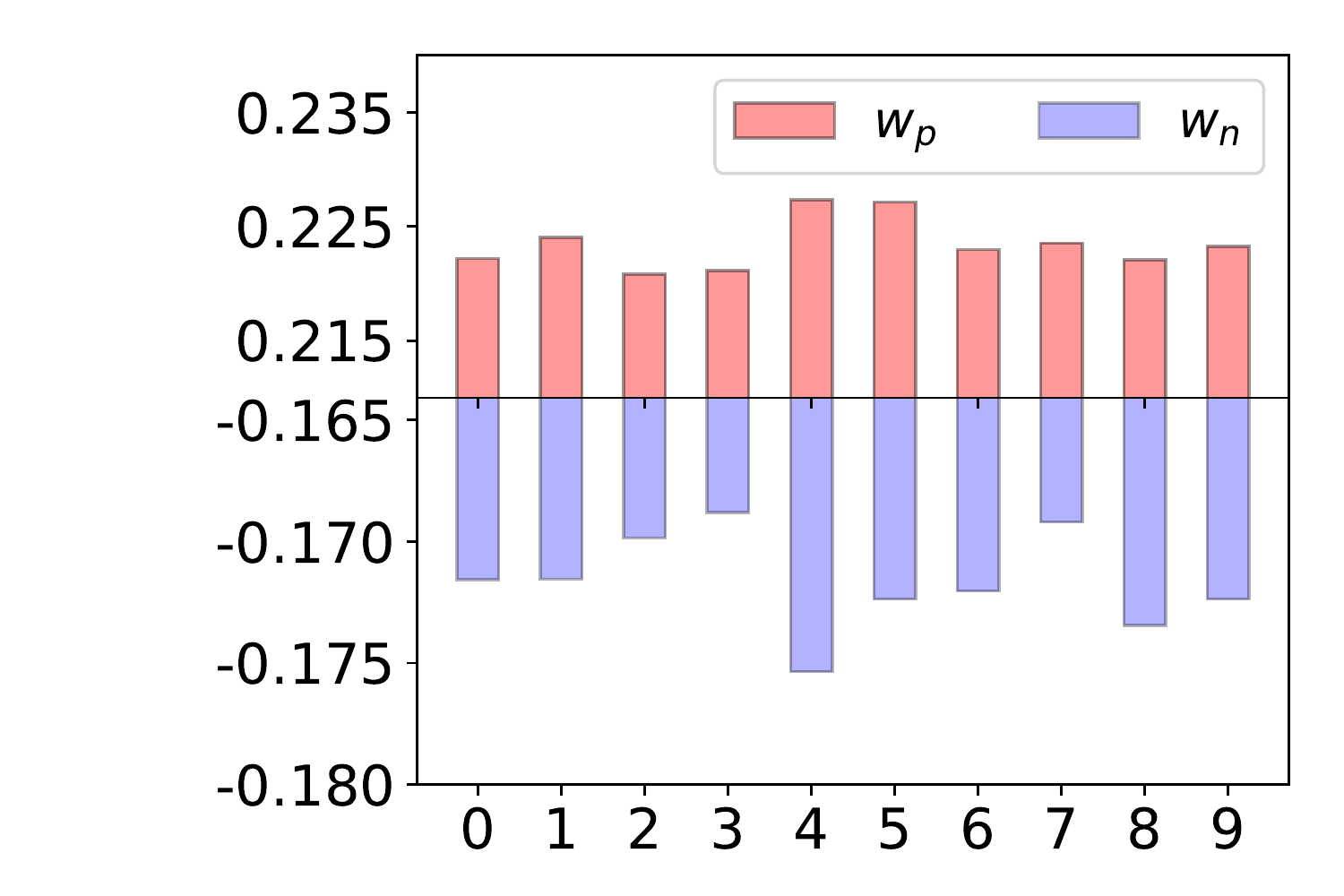}
	
\caption{Values of $w_p$ and $w_n$ in 10 independent experiments.}\label{fig.para}
\end{figure}

\subsection{Parameter Analysis} 
In this section, we conducted several studies on the results of other two key parameters learned by our model, i.e., $w_p$ and $w_n$.
The results in 10 independent experiments are illustrated in Fig.~\ref{fig.para}.
According to these results, we find $w_p$ is consistently positive.
It is intuitive because the news read by users carefully can usually represent the preferences of users.
Thus, the similarities between the candidate news and the positive clicked news should have a positive impact on the click probability.
In addition, we find $w_n$ is consistently negative. 
This may be because if a user reads a news article very quickly, we can infer that he/she is probably not interested in this news.
Thus, the click probability should have a negative correlation with the similarities between candidate news and negative clicked news.

\begin{figure}[t]
	\centering
	\includegraphics[width=0.55\textwidth]{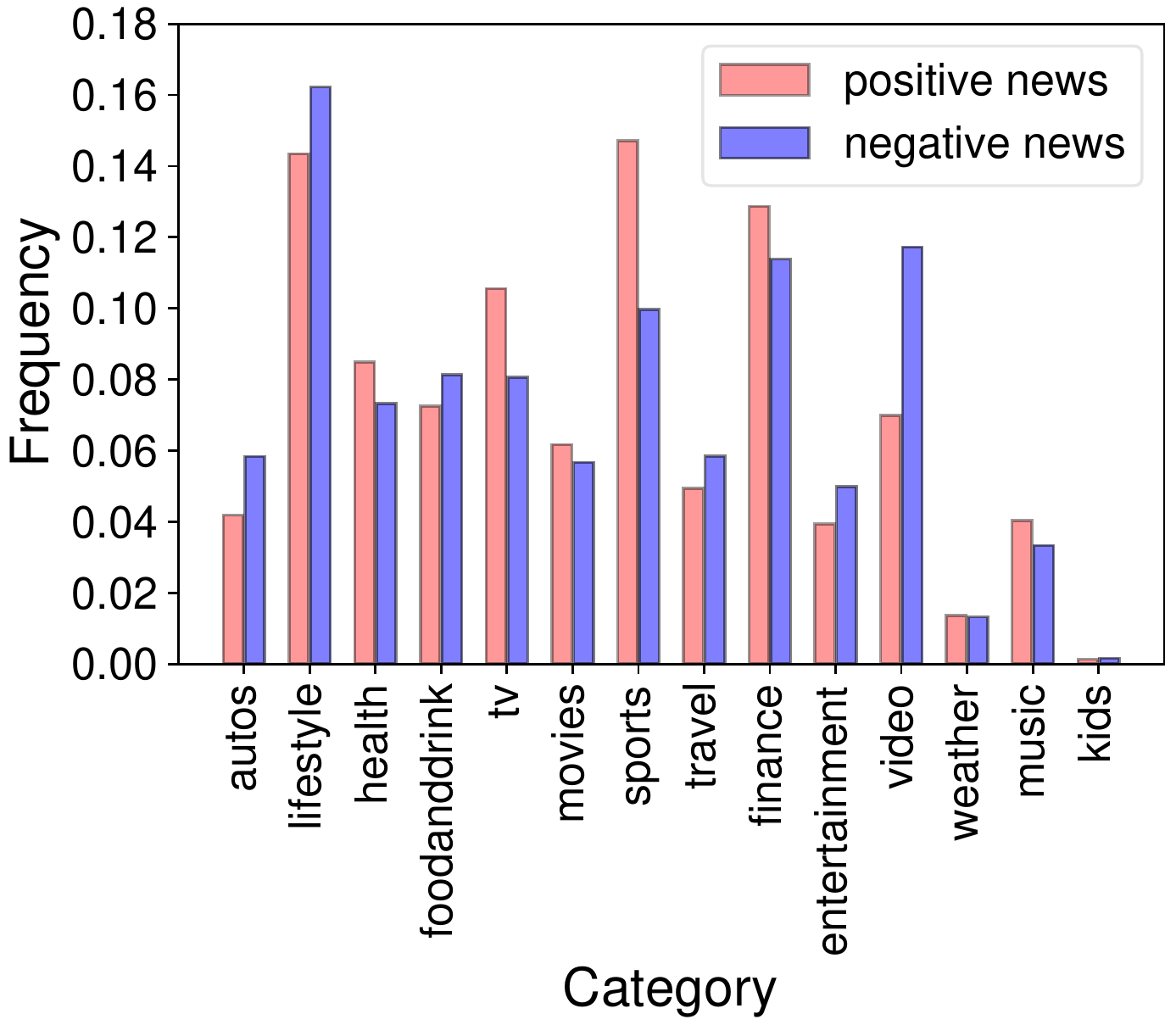}
\caption{Topic distributions of positive and negative clicked news.}\label{fig.topic}
\end{figure}

\subsection{Qualitative Analysis of Negative News}
In this section, we will conduct several qualitative analysis on the characteristics of the negative clicked news.
First, we want to compare the topic distributions of positive and negative news, and the results are shown in Fig.~\ref{fig.topic}.
According to Fig.~\ref{fig.topic}, we find the ratios of negative news in different topic categories have some differences.
For example, it is interesting that the ratio of negative news is the highest in the ``video'' category.
This may be because many users are attracted by the titles of the news in this category,  but they are not interested in the content of these news.
Thus, incorporating the information of negative feedback may also implicitly encode topical information into user representations, which is useful for more accurate news recommendation.

\begin{table}[t]
\caption{Top News with the highest negative feedback ratios.}\label{table.nf}
\begin{tabular}{|l|c|}
\hline
\multicolumn{1}{|c|}{Clicked News}                                                    &NF Ratio \\ \hline
The 30 Most Heartwarming Random Acts of Kindness from 2018                                    & 75.8\%   \\
Pink Lettuce Is the Newest Pink Produce                             & 73.8\%    \\
10 Secrets of Airline Food                        & 73.4\% \\
Get Ready to Load Up Your Plates! 20 Restaurants Where Kids Eat Free                      & 73.2\%  \\
Christmas coat drive for Paterson students                              & 73.0\%  \\ 
The Best All-You-Can-Eat Deal in Every State                     & 72.0\%  \\
The Most Ingenious Foods Featured On Shark Tank                      & 71.7\%  \\
The Five Bikes Sabrina is Most Excited to See in 2019                      & 70.3\%  \\
The greatest sports siblings in history                    & 67.1\%  \\
2018 Holiday Gift Guide: The 25 Best Gift Ideas Under \$500                     & 66.7\%  \\
\hline
\end{tabular}
\end{table}

In addition, we aim to discover some common patterns of negative news, e.g., the contents and writing styles of their titles.
We calculate the ratio of negative feedback (NF) for each news\footnote{We ignore the news with less than 10 clicks to filter some possible noisy news.}, and the titles of the top 10 news with the highest NF ratios are listed in Table~\ref{table.nf}.
From Table~\ref{table.nf}, we find many news with top NF ratios are clickbaits, and they have very similar writing styles, i.e., describing something in a sensationalized way but no details are provided.
Thus, many users may be attracted by the news titles and click these news.
However, after taking a glance at the news body, many users may be disappointed and close these news quickly.
Thus, it is harmful for news platforms to improve user experiences if clickbaits are frequently recommended to users.
Fortunately, our approach may have the potential to  avoid recommending too many clickbaits to user by incorporating the negative feedback of users, which can improve the reading experiences of users.

\begin{table*}[t]
\caption{Several clicked and candidate news of a randomly selected user. The maximum predicted score in each column and the candidate news clicked by this user are in bold. NF represents negative feedback.}\label{table.case}
\resizebox{1\textwidth}{!}{
\begin{tabular}{|l|c|c|}
\hline
\multicolumn{1}{|c|}{Clicked News}                                                                            & \multicolumn{2}{c|}{Dwell time} \\ \hline
23 Celebs Who Still Live in Their Hometown                                           & \multicolumn{2}{c|}{75s (+)}    \\
Ranking the Oscar's Best Picture winners from every year                             & \multicolumn{2}{c|}{1s (-)}     \\
Oscars 2019: 'The Favourite,' 'Roma' Lead With 10 Nominations                        & \multicolumn{2}{c|}{2s (-)}     \\
Gladys Knight to Perform the National Anthem at Super Bowl LIII                      & \multicolumn{2}{c|}{44s (+)}    \\
Five Players Who Could Be Traded During Super Bowl Week                              & \multicolumn{2}{c|}{187s (+)}   \\ \hline
\multicolumn{1}{|c|}{Candidate News}                                                 & Score (w/o NF) & Score (w/ NF)  \\ \hline
\multicolumn{1}{|l|}{\textbf{Ranking the possible Super Bowl LIII matchups}}         & 0.669          & \textbf{0.893} \\
Bradley Cooper Snubbed by Oscars for Best Director And More Nomination Shocker       & \textbf{0.912} & 0.075          \\
\multicolumn{1}{|l|}{Olympic swimmer Nathan Adrian reveals he has testicular cancer} & 0.212          & 0.158          \\ \hline
\end{tabular}
}
\end{table*}
\subsection{Case Study}
In this section, we conducted several case studies to show the effectiveness of our approach.
First, we want to visually explore the effectiveness of incorporating the negative feedback of users.
The clicked news and candidate news of a randomly selected user are listed in Table~\ref{table.case}.
We calculate the click probability score respectively using our approach and its variant without negative feedback.
According to Table~\ref{table.case}, the second candidate news is assigned the highest click score when negative feedback is not considered.
This is probably because it is very similar to the second and third clicked news, since all of them are about Oscars. 
However, we can infer that this user is probably not very interested in this topic since she closes the second and third clicked news very quickly.
Thus, it may be ineffective to recommend Oscars related news to this user.
Fortunately, our approach can assign the second candidate news a low click score.
This is because our approach can  take the implicit negative user feedback into consideration, which is beneficial for modeling user preferences more accurately.

\section{Conclusion}\label{sec:Conclusion}

In this paper, we propose a neural news recommendation approach that can consider the implicit negative feedback of users.
We propose to distinguish positive and negative news clicks according to users' reading dwell time, and we propose to respectively learn two different embeddings of a user from her positive and negative news clicks.
In addition, we propose to compute a unified news click score based on a combination of the relevance scores between candidate news and these two user embeddings.
Besides, we propose an interactive news modeling method that can learn unified news representations from title and body with full consideration of their relatedness.
Extensive experiments on real-world dataset show that incorporating negative user feedback can effectively improve the performance of user modeling for news recommendation.

\begin{acknowledgements}
The authors would like to thank Microsoft News for providing technical support and data.
This work was supported by the National Key Research and Development Program of China under Grant number 2018YFC1604002, and the National Natural Science Foundation of China under Grant numbers U1936216, U1936208, U1705261 and U1836204.
\end{acknowledgements}

\section*{Conflict of interest}
The authors declare that they have no conflict of interest.

\bibliographystyle{spmpsci}      
\bibliography{weibo}   

\end{document}